\documentclass[aps,prd,preprint,superscriptaddress]{revtex4-2}
\usepackage{graphicx}
\graphicspath{{img/}}
\usepackage[makeroom]{cancel}
\usepackage{xcolor}
\usepackage{float}
\usepackage{amsmath,amssymb,amsfonts,amsbsy}
\usepackage{hyperref}  

\newcommand{\GeV}{\mathrm{GeV}}

\begin{document}

\title{Axial meson contribution to proton form factors}


\author{N.Korchagin}
 \email{korchagin@impcas.ac.cn}
 \affiliation{Institute of Modern Physics, Chinese Academy of Sciences, Lanzhou 730000, People Republic of China}

\author{A.Radzhabov}
\email{aradzh@icc.ru}
\affiliation{Matrosov Institute for System Dynamics and Control Theory SB RAS, 664033, Irkutsk, Russia}
\affiliation{Irkutsk State University, 664003, Irkutsk,  Russia}




\begin{abstract}
The effect of axial meson exchange on extraction of proton electromagnetic form factors in $ep$ scattering is calculated.
Taking as an example $f_1(1285)$ meson exchange, it is quantitatively shown how it affects Rosenbluth and polarization transfer methods.
The contribution of an axial meson is negligible in comparison with the  traditional box-type two photon exchange and could not be responsible for difference of form factors extracted by different techniques. However, results support the idea that meson exchange could explain deviation of individual polarizations.
\end{abstract}



\maketitle


\section{Introduction}


Understanding of the internal structure of proton is a cornerstone of the theory of strong interaction.
Effectively the proton composition is encoded in form factors.
In electron-proton scattering one can measure electric $G_E$ and magnetic $G_M$ form factors (also called as Sachs form factors).
The first is related with electric charge distribution and the second is with magnetization.


There are two techniques to extract $G_{E,M}$. Both utilize the one photon exchange(OPE) approximation.
First method uses unpolarized $ep$ cross section measurements.
It exploits the linearity of the cross-section in OPE approximation with respect to some kinematic variable $\varepsilon$(meaning will be discussed later) to extract $G_{E,M}$.
This method is known as Rosenbluth technique(proposed by Rosenbluth \cite{Rosenbluth:1950yq}).


Second method is used in experiments with polarized beams.
It was proposed by Akhiezer\cite{AkhiezerJETP:1958} and later developed in works \cite{Akhiezer:1968ek,Akhiezer:1974em}.
Longitudinally polarized electron is scattered off the unpolarized nucleon and the recoil nucleon gets a polarization $\vec{e}+N \to e + \vec{N}$.
The ratio of transverse to longitudinal polarizations of the recoil nucleon is proportional to $G_E/G_M$.
This technique is also called polarization transfer method.


Surprisingly behavior of form factors obtained by different methods disagrees.
In Rosenbluth technique $G_E$ and $G_M$ fall with the same rate with $Q^2$.
The ratio $\mu_p G_E/G_M$ is close to 1, while from polarization transfer experiments this ratio decreases with $Q^2$.
For more details we refer the reader to reviews \cite{PerdrisatReview:2006,PacettiReview:2015,PunjabiReview:2015}.


It is common belief that discrepancy originates from higher $\alpha$ order corrections.
They can be divided on real(bremsstrahlung) and virtual: vacuum polarization, self-energy corrections, vertex corrections and two photon exchange(TPE).
For polarization transfer bremsstrahlung is small in comparison with TPE effects \cite{BorisyukKobushkin:2014}.
Due to the nature of measuring the ratio of cross sections, corrections are almost canceled.
For the available $Q^2$ range bremsstrahlung gives less than $1\%$ correction to $G_E/G_M$.
Meanwhile TPE contribution grows significantly at high $Q^2$, especially at high beam energy.
On the other hand, for Rosenbluth method, bremsstrahlung is significant and it is considered during data analysis.
However there are concerns about its completeness: additional higher order corrections and large logarithms which are not included to analysis could be important \cite{Arbuzov:2015vba,Arbuzov:2019blj}.

\begin{figure}
\centering
\includegraphics[scale=0.58]{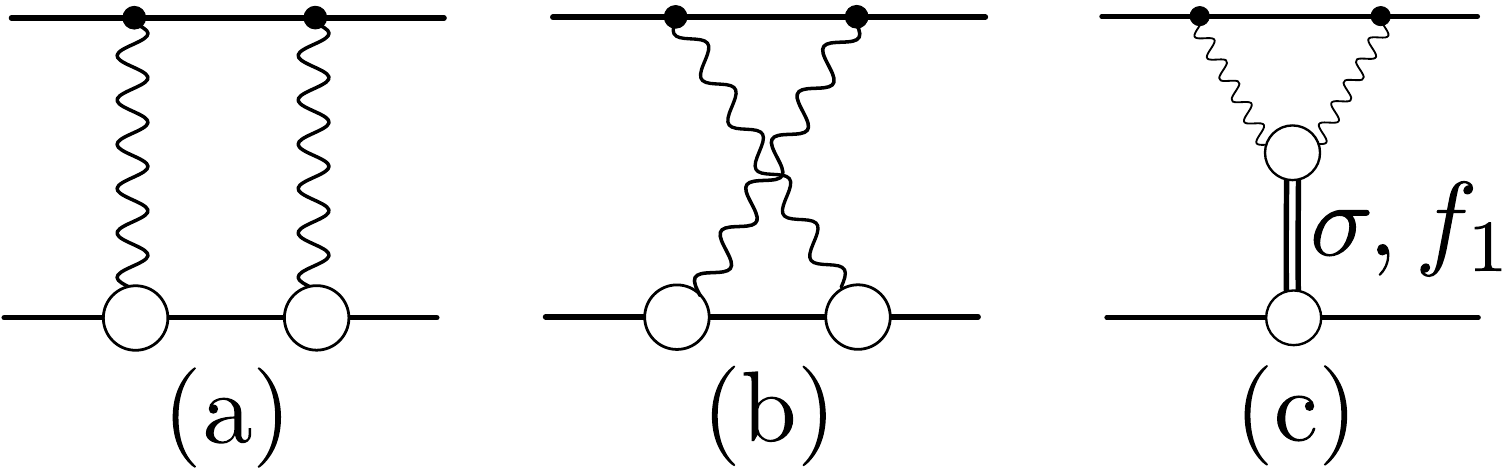}
\caption{Possible TPE diagrams. Box, cross-box and the meson exchange diagrams. Blobs represent non-point-like vertices.}
\label{fig:TPE}
\end{figure}

Commonly TPE is represented as box and crossed-box diagrams,
Fig.\ref{fig:TPE}(a,b).
Contributions with elastic and inelastic intermediate states could resolve inconsistency between form factors extracted by Rosenbluth and polarization transfer methods \cite{BlundenMelnitchoukTjon:2005}.
However, other polarization observables tend to disagree with most theoretical models. For example  measurement of the longitudinal recoil polarization $P_L$ has noticeable deviation at forward scattering angles from OPE and TPE corrections worsen consistency\cite{GEp2gammaCollMeziane:2010,Puckett:2017flj}.
Moreover, only part of such calculation for TPE can be done in a model independent way.
A structure of intermediate hadron states influence results(for more details see reviews \cite{ArringtonReview:2011,AfanasevTPEreview:2017}).

There is an additional TPE contribution from exchange of a single $C$-even meson, which is connected to the lepton by two photons, Fig.\ref{fig:TPE}(c).
Such diagrams were examined in \cite{ChenZhou:2013, KoshchiiAfanasev:2016}.
Pseudoscalar meson contribution is exactly zero for the cross section of unpolarized particles.
Authors of \cite{KoshchiiAfanasev:2016} are focused at MUSE experiment kinematics, i.e. small beam energies and $Q^2$. In such case the primary candidate for the exchanged particle is $\sigma$-meson.
Here we concentrate on the high $Q^2$ kinematics region where the  deviation appears between cross section and polarized data.


The contribution of axial-vector mesons was proposed long time ago by Drell and Sullivan \cite{Drell:1965is} as a possible explanation for the deviation of theory from experiment for the hyperfine splitting.
In the recent paper \cite{Dorokhov:2017nzk} contribution of axial mesons in muonic hydrogen spectroscopy was calculated. The effect is significant for the Zemach radii although small for the charge radius.
Most important is that the meson correction is the same order as TPE for $1S$ level.
This fact inspired us to calculate the effect of axial mesons on extraction of form factors from elastic $ep$ scattering.
We repeat analytical results of \cite{RekaloTomasi:2004} and also estimate quantitatively axial-vector contribution.


This work is organized in the following way.
In the next section the formalism of OPE approximation is briefly discussed.
In the section \ref{sec:axial_contr} the contribution of axial vector exchange to polarization transfer and cross section is calculated.
Then, the section \ref{sec:couplings} discusses the calculation of effective electron-meson and proton meson couplings.
The last section \ref{sec:results} presents the numerical results for the  new contribution.

\section{Elastic lepton-proton scattering formalism}\label{sec:breit_frame}

\begin{figure}[h]
\centering
\includegraphics[scale=0.58]{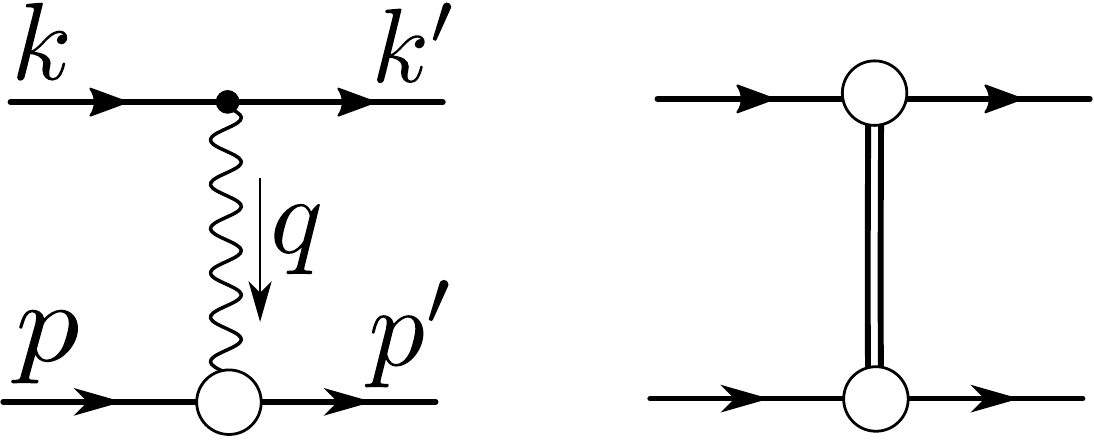}
\caption{One photon and meson exchange contributions. Blobs represent non point-like vertices.}
\label{fig:vec_and_meson_tree_diag}
\end{figure}

We consider elastic lepton-proton scattering  $l(k) + p(p) \to l(k') + p(p')$.
The exchanged photon or meson has momenta
\begin{equation}
  q=p'-p=k-k', \qquad q^2=-Q^2.
\end{equation}
The full amplitude $\mathcal{M}$ is the sum of the photon exchange $\mathcal{M}_{\gamma}$ and axial meson exchange $\mathcal{M}_{A}$ amplitudes
\begin{equation}
  \mathcal{M}^2=|\mathcal{M}_{\gamma}+\mathcal{M}_{A}|^2 \approx |\mathcal{M}_{\gamma}|^2 + 2\text{Re} \mathcal{M}_{\gamma}\mathcal{M}_{A}^*.
\end{equation}
$|\mathcal{M}_{A}|^2$ is suppressed by a higher orders of $\alpha$.
$\mathcal{M}_{\gamma}$ and $\mathcal{M}_{A}$ are represented by the left and right Feynman diagrams in Fig.~\ref{fig:vec_and_meson_tree_diag} respectively.
The meson-lepton coupling is treated as an effective axial vertex with the coupling $g_{eA}$. 
This coupling is calculated in the section \ref{sec:couplings}.

Amplitudes are expressed in terms of lepton and hadron currents. Vector currents are
\begin{align}
  j^{(\gamma)}_{\mu} &= Q_l e \bar{u}(k')\gamma_{\mu}u(k), \\
  J^{(\gamma)}_{\mu} &= Q_h e \bar{U}(p')\Big[\gamma_{\mu}F_1(Q^2) - \frac{\sigma_{\mu \nu} q_{\nu}}{2 M_p} F_2(Q^2)\Big] U(p),
\end{align}
where $Q_l, Q_h$ are lepton and nucleon charges in units of electron charge(i.e. $-1$ for electron, $+1$ for proton), $e$ is the electron charge($e^2/4\pi=\alpha_{QED}$),
$\sigma_{\mu\nu} = \frac{1}{2} [\gamma_{\mu}, \gamma_{\nu} ]$, $M_p$ is the proton mass.
$F_1$ and $F_2$ are Dirac and Pauli form factors.

For the axial meson exchange the lepton and hadron axial currents are
\begin{align}
  j^{(A)}_{\mu} &= g_{eA}(Q^2) \bar{u}(k') \gamma_{\mu} \gamma_5 u(k), \\
  J^{(A)}_{\mu} &= g_{pA}(Q^2)\bar{U}(p')\gamma_{\mu} \gamma_5 U(p),
\end{align}
where $g_{eA}(Q^2)$ is the effective coupling of a lepton with an axial meson and $g_{pA}(Q^2)$ is the proton-meson coupling.
Further the $Q^2$ dependency of couplings is omitted for shortness.
\begin{figure}[tbh]
\centering
\includegraphics[scale=0.8]{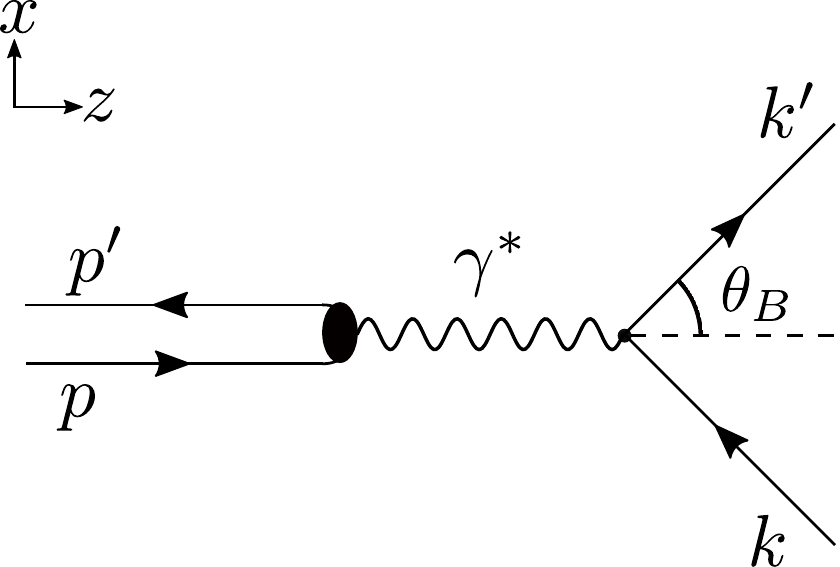}
\caption{Breit frame kinematics.}
\label{fig:breit_frame}
\end{figure}

It is convenient to do calculation in Breit frame (reader can find pedagogical introduction in this technique in lecture notes \cite{RekaloLecture:2002gv}).
This frame is chosen in such a way that the photon energy is $0$. 
The photon 3-momentum is chosen along $+z$ direction $\vec{q}=(0,0,q_B)$,
scattering is in the $xz$-plane and the electron mass is neglected. Momenta in this frame are
\begin{align}
k&=(E_e, \frac{q_B}{2} \cot \frac{\theta_B}{2},0,\frac{q_B}{2}), & k'&=(E_e,\frac{q_B}{2} \cot \frac{\theta_B}{2},0,-\frac{q_B}{2}), &  E_e^2= \frac{Q^2}{4\sin^2 \theta_B/2}, \\
p&=(E_p,0,0,-q_B/2), &  p'&=(E_p,0,0,q_B/2), & E_p^2= M_p^2(1+\tau),
\end{align}
where $\tau = Q^2/4M_p^2$ and $\theta_B$ is the electron scattering angle in Breit frame. 
The $\theta_B$ and the Lab frame scattering angle $\theta_e$ are related as
\begin{equation}
  \cot^2(\theta_B/2)=\frac{\cot^2(\theta_e/2)}{1+\tau}.
\end{equation}
It is convenient to introduce a following dimensionless variable $\nu$:
\begin{equation}
\nu = \frac{p\cdot k}{M_p^2} - \tau = \frac{s-u}{4M_p^2} = \frac{E_e E_p}{M_p^2} = \frac{E_1+E_2}{2M_p} = \sqrt{\tau(1+\tau)} \sqrt{\frac{1+\varepsilon}{1-\varepsilon}},
\end{equation}
where $E_{1,2}$ are initial and final electron energies in Lab frame
\begin{equation}
E_1=M_p(\nu+\tau), \qquad E_2 = E_1-2M_p \tau = M_p(\nu-\tau),
\end{equation}
and $\varepsilon$ is
\begin{equation}
\varepsilon = \frac{1}{1+2(1+\tau)\tan^2(\theta_e/2)}.
\end{equation}
%
The $\varepsilon$ is often erroneously called in literature as a degree of longitudinal polarization. In fact, it is a degree of linear polarization \cite{Akhiezer:1974em}.

The contribution of meson exchange in unpolarized and polarized scattering is calculated in the next section in the described framework.

\section{Axial vector contribution}\label{sec:axial_contr}

The OPE amplitude is well known and it has a simple form in Breit frame \cite{RekaloLecture:2002gv}:
\begin{align}\label{photon_amp}
\overline{|\mathcal{M}_{\gamma}|}^2 & = \frac{1}{4}\sum_{\text{spin}} j^{(\gamma)}_{\mu} (j^{(\gamma)}_{\mu'})^* J^{(\gamma)}_{\nu} (J^{(\gamma)}_{\nu'})^* \frac{g^{\mu\nu} g^{\mu'\nu'}}{Q^4}  \\
&= (Q_l^2 Q_h^2) e^4 \frac{4 M_p^2}{Q^2} \Big((G_E^2 + \tau G_M^2) \cot^2(\theta_B/2) + 2 \tau G_M^2 \Big) \\
&=(Q_l^2 Q_h^2) e^4 \frac{1}{\tau(1+\tau)} \Big(G_E^2 + \frac{\tau}{\varepsilon} G_M^2 \Big) \cot^2(\theta_e/2).
\end{align}
The last line corresponds to transition to the Lab frame.
Electric and magnetic form factors $G_{E,M}$ are related with Dirac and Pauli form factors as
\begin{equation}\label{EMFF}
G_E=F_1-\tau F_2, \qquad G_M=F_1+F_2.
\end{equation}
%


Calculation for interference between OPE and meson exchange amplitudes is straightforward and gives
\begin{align}\label{eq:int_un_FC}
    2 \text{Re} \mathcal{M}_{\gamma}\mathcal{M}_{A}^{\dagger} &=
        2 \frac{1}{4} \sum_{\textrm{spin}} j_{\mu}^{(A)} (j_{\mu'}^{(\gamma)})^*
        J_{\nu}^{(A)} (J_{\nu'}^{(\gamma)})^*
        \frac{g_{\mu \nu}}{q^2} \frac{g_{\mu' \nu'} - q_{\mu'}q_{\nu'}/q^2}{q^2 - m_A^2} \\
&= 2 (Q_l Q_h e^2)\frac{g_{eA} g_{pA}}{(Q^2+m_A^2)Q^2} L^{(A\gamma)}_{\mu\nu} W^{(A\gamma)}_{\mu\nu} \\
&= (Q_l Q_h e^2)\frac{g_{eA} g_{pA}}{(Q^2+m_A^2)Q^2} 16 E_e E_p Q^2 G_{M} \\
&= (Q_l Q_h e^2)\frac{g_{eA} g_{pA}}{(Q^2+m_A^2)} \Big( \frac{8 M_p Q}{\sin(\theta_B/2)} \sqrt{(1+\tau)} G_{M} \Big),
\end{align}
where $m_A$ is the mass of an axial meson. 
Notice that $L^{(A\gamma)}_{\mu\nu} W^{(A\gamma)}_{\mu\nu}$ is spin averaged and does not contain couplings.

The cross section is related with the amplitude in Lab frame as
\begin{equation}\label{eq:cross_section}
    \frac{d\sigma}{d\Omega} = \frac{|\mathcal{M}|^2}{64 \pi^2}
    \left( \frac{E_2}{E_1} \right)^2 \frac{1}{M_p^2}.
\end{equation}
Historically the $ep$ cross section is represented as a correction to the Mott cross section, which is a cross section for scattering of spin $1/2$ point-like particles
\begin{equation}\label{eq:mott}
  \frac{d\sigma_M}{d\Omega} = \frac{\alpha^2}{Q^2}
    \left(
        \frac{E_2}{E_1}\right)^2 \cot^2 \left(\frac{\theta_e}{2}
    \right).
\end{equation}
The final formula for cross-section is
\begin{equation}\label{eq:AV_correction_to_xsection}
    \frac{d\sigma}{d\Omega} = \frac{d\sigma_M/d\Omega}{\varepsilon(1+\tau)}
    \Bigg(
        \varepsilon G_E^2 + \tau G_M^2 + \frac{(Q_l Q_h)g_{eA} g_{pA}}{2 \alpha \pi}
        \frac{\tau \nu (1-\varepsilon)}{\tau + m_A^2/4M_p^2} G_M
    \Bigg).
\end{equation}
The part in parenthesis is so-called reduced cross section. The last term  corresponds to the meson exchange contribution. 
It behaves as $\sqrt{1-\varepsilon^2}$ in comparison with the OPE part $(\varepsilon G_E^2+\tau G_M^2)$.
At $\varepsilon \to 1$, which corresponds to forward scattering, the interference correction goes to $0$. 
At $\varepsilon\to 0$ correction is maximum.
Our result is similar to the old calculation of Drell and Sullivan \cite{Drell:1965is} but differs by factor 2 from Eq.~(8) in~\cite{RekaloTomasi:2004}.


As the next step the meson exchange contribution in polarized scattering is calculated. The setup is following: there is scattering of a massless, longitudinally polarized electron $k$ with helicity $\lambda$ on the unpolarized proton $p$.
The recoil proton $p'$ receives polarization, described by the vector $a^{\mu}$.
In the OPE approximation, the product of lepton and hadron tensors in Breit frame is
\begin{align}
L^{(\gamma\gamma)}_{\mu\nu}W^{(\gamma\gamma)}_{\mu\nu}(a) &=\frac{1}{4}\text{Tr}[\cancel{k}'\gamma_{\mu} \cancel{k} (\lambda \gamma_5) \gamma_{\nu}] \nonumber \\
&\times
\text{Tr}[(\cancel{p}'+M_p)(1-\gamma_5 \cancel{a})\Big(\gamma_{\mu}F_1 - \frac{\sigma_{\mu\rho}q_{\rho}}{2M_p}F_2 \Big)(\cancel{p}+M_p)  \Big(\gamma_{\nu}F_1 - \frac{\sigma_{\rho \nu}q_{\rho}}{2M_p}F_2 \Big)] \nonumber \\
&=
\frac{16 \lambda M_p \tau}{1+\tau} G_M \Big[ E_e E_p (a \cdot q) \big(G_E-G_M\big) + M_p^2 (1+\tau) G_E (a \cdot (k+k')) \Big],
\end{align}
where summation over the polarization of the final lepton and average over the polarization of the initial proton is preformed.
The 4-vector $a^{\mu}$ is related with the rest frame polarization vector $\xi^{\mu}=(0,\vec{\xi})$ by boost:
\begin{equation}
  a_{\mu}=\Big( \frac{\vec{\xi} \cdot \vec{p}}{M_p},\vec{\xi}+\frac{(\vec{\xi} \cdot \vec{p}) \vec{p} }{M_p(E_p+M_p)} \Big).
\end{equation}
The products $(a\cdot q)$ and $(a\cdot (k+k'))$ for different polarizations in Breit frame are
\begin{align}\label{eq:scalar_products_for_xyz_polarizations}
  &P_x:&  &(a_x\cdot q) = 0,  &(a_x \cdot (k+k')) &= - q_B \cot(\theta_B / 2 ), \nonumber \\
  &P_z:&  &(a_z\cdot q) = -\frac{q_B}{M_p} E_p, &(a_z \cdot (k+k')) &= \frac{q_B}{M_p}E_e, \\
  &P_y:&  &(a_y\cdot q) = 0, &(a_y \cdot (k+k')) &= 0. \nonumber
\end{align}
%


For $P_x$ and $P_z$ polarizations 
the well known formulas \cite{Akhiezer:1968ek,Akhiezer:1974em} are reproduced
\begin{align}
    L^{(\gamma\gamma)}_{\mu\nu}W^{(\gamma\gamma)}_{\mu\nu}(P_x)
        &=-4 \lambda M_p Q^3 G_M G_E \cot(\theta_B/2)
        =-32\lambda M_p^4 \tau \sqrt{\frac{2\varepsilon\tau}{1-\varepsilon}} G_E G_M,
\\
    L^{(\gamma\gamma)}_{\mu\nu}W^{(\gamma\gamma)}_{\mu\nu}(P_z)
        &= 2 \lambda \frac{Q^4 G_M^2}{\sin(\theta_B/2)}
        = 32 \lambda M_p^4 \tau^2 \sqrt{\frac{1+\varepsilon}{1-\varepsilon}}G_M^2.
\end{align}
For $P_y$ polarization the amplitude is zero in OPE approximation.


The interference between photon and meson exchange amplitudes is
\begin{equation}
  L^{(A\gamma)}_{\mu\nu}W^{(A\gamma)}_{\mu\nu}(a)= 16 \lambda M_p  \Big[(a \cdot q) \big((E_e^2+M_p^2 \tau)G_M - E_e^2G_E\big) -E_e E_p G_E (a \cdot (k+k')) \Big].
\end{equation}
It is worthwhile to note the absence of $g_{pA}$ because it is the common factor ``outside'' $L^{(A\gamma)}_{\mu\nu}W^{(A\gamma)}_{\mu\nu}$ in Eq.~(\ref{eq:int_un_FC}).
Using Eq.~(\ref{eq:scalar_products_for_xyz_polarizations}) for longitudinal and transverse polarization one can get
\begin{align}
  L^{(A\gamma)}_{\mu\nu}W^{(A\gamma)}_{\mu\nu}(P_x) &= -8 \lambda Q^2 M_p^2 \frac{\sqrt{1+\tau}}{\sin(\theta_B/2)} \cot(\theta_B/2) G_E \\
  &= -8 \lambda Q^2 M_p^2 \frac{\sqrt{2 \varepsilon (1+\varepsilon) (1+\tau)}}{1-\varepsilon} G_E,
  \\
  L^{(A\gamma)}_{\mu\nu}W^{(A\gamma)}_{\mu\nu}(P_z) &= 4 \lambda M_p Q^3 \sqrt{1+\tau} (1 + \cot^2(\theta_B/2)) G_M
  \\
  &= 8 \lambda M_p Q^3 \frac{\sqrt{1+\tau}}{1-\varepsilon}G_M.
\end{align}
The interference does not give contribution to $P_y$ polarization because $L^{(A\gamma)}_{\mu\nu}W^{(A\gamma)}_{\mu\nu}(P)$ depends on the same scalar products as in OPE. Also only imaginary part of TPE gives rise of $P_y$.

The final result for polarization transfer is
\begin{align}\label{eq:AV_polarization_correction_result}
   \frac{d \sigma}{d \Omega}(P_x) &=
   -\lambda \frac{\sigma_M / d\Omega}{\varepsilon (1+\tau)} \sqrt{2 \varepsilon (1-\varepsilon)\tau}
   \Big[ G_E G_M + (Q_l Q_h) \frac{g_{eA} g_{pA} G_E}{4 \alpha \pi} \frac{\nu}{\tau+m_A^2/4M_p^2} \Big], \\
    \frac{d \sigma}{d \Omega}(P_z) &= \lambda \frac{\sigma_M / d\Omega}{\varepsilon (1+\tau)} \tau \sqrt{1-\varepsilon^2} \Big[ G_M^2 + (Q_l Q_h)\frac{g_{eA} g_{pA} G_M}{2\alpha\pi(1+\varepsilon)} \frac{\nu}{(\tau+m_A^2/4M_p^2)} \Big].
\end{align}
%
One can rewrite it in the form as in \cite{RekaloTomasi:2004} and 
result for $P_x$ polarization has an additional factor $\sqrt{\frac{1+\tau}{\tau}}$.


The all contributions to the cross section and polarization transfer is obtained.
To get numerical result and compare with experiments it is necessary  
to know the effective couplings $g_{eA}$ and $g_{pA}$.  
In the next section
that couplings are  estimated.


\section{Effective couplings}\label{sec:couplings}

This section is dedicated to calculation of the effective couplings of an axial meson to the electron $g_{eA}$ and to the proton $g_{pA}$.
$f_1(1285)$ is taken as an example of axial meson.


\begin{figure}[tbh]
	\includegraphics[scale=0.8]{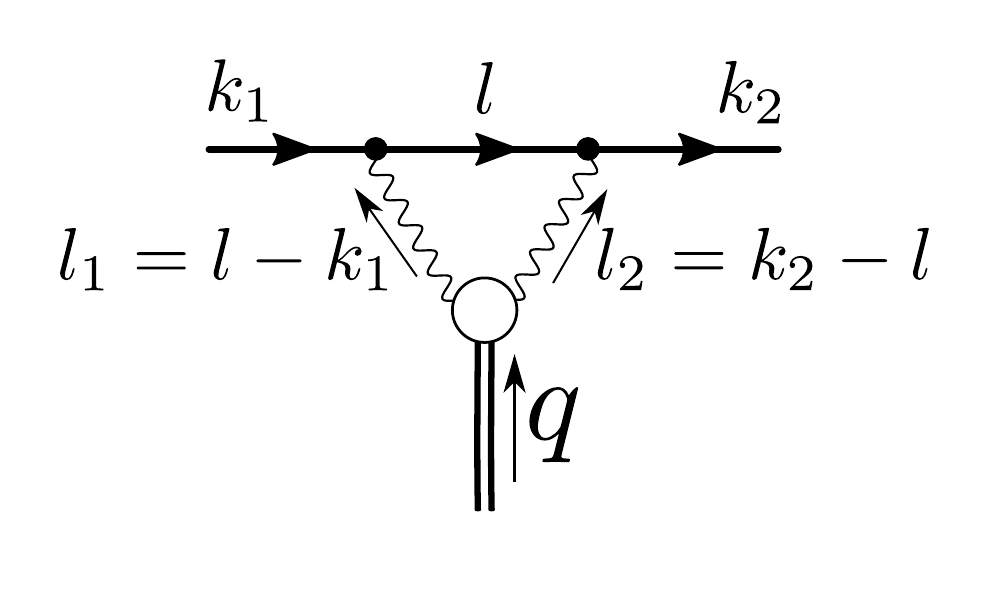}
	\caption{\label{fig:loop} The effective lepton-meson coupling is given by the triangle loop.}
\end{figure}

The meson-photons vertex can be parameterized in the form\cite{Pascalutsa:2012pr}:
\begin{align}\label{eq:VAV-vertex}
\mathcal{T}^{\mu\nu}_\alpha =& \,  i e^2 \varepsilon_{\rho \sigma \tau \alpha} \,  \Bigg\{
R^{\mu \rho} (l_1, l_2) R^{\nu \sigma} (l_1, l_2) \,
(l_1 - l_2)^\tau \, \nu \, F^{(0)}(l_1^2, l_2^2)
 \nonumber \\
& + \, R^{\nu \rho}(l_1, l_2) \left( l_1^\mu - \frac{l_1^2}{\nu} l_2^{\mu} \right)
l_1^\sigma \, l_2^\tau \, F^{(1)}(l_1^2, l_2^2) \nonumber \\
& + \, R^{\mu \rho}(l_1, l_2)
\left( l_2^\nu - \frac{l_2^2}{\nu} l_1^{\nu} \right)
l_2^\sigma \, l_1^\tau \, {F^{(1)}}(l_2^2, l_1^2)
\Bigg\},
\end{align}
where superscripts in form factors $F^{(0)}$ and $F^{(1)}$ indicate a helicity state of an axial meson.
$F^{(0)}$ is symmetric under $l_1 \leftrightarrow l_2$.
Normalization factor $1/m_A^2$ which appears in \cite{Pascalutsa:2012pr} is included to form factors.
Note that this equation is for the case when a meson is on the mass-shell.
The off-shellness is considered as additional $q^2$ dependency of $F^{(i)}$ form factors and discussed further.
Other ingredients of the Eq.~(\ref{eq:VAV-vertex}) are
\begin{align}
\nu &= (l_1 \cdot l_2),\\
X &= (l_1 \cdot l_2)^2 - l_1^2 l_2^2 , \\
R^{\mu \nu} (l_1, l_2) &= - g^{\mu \nu} + \frac{1}{X} \,
\bigl \{
\nu \left( l_1^\mu \, l_2^\nu + l_2^\mu \, l_1^\nu \right)
- l_1^2 \, l_2^\mu \, l_2^\nu  - l_2^2 \, l_1^\mu \, l_1^\nu
\bigr \} ,\\
l_{1 \mu} R^{\mu \nu} (l_1, l_2) &= l_{2 \mu} R^{\mu \nu} (l_1, l_2)= l_{1 \nu} R^{\mu \nu} (l_1, l_2)= l_{2 \nu} R^{\mu \nu} (l_1, l_2) =0.
\end{align}
%


The effective electron-meson coupling is determined by the loop integral, Fig.~\ref{fig:loop}
\begin{align}
\bar{u}(k_2) \Big[ e^2 \int \frac{d^4 l}{(2\pi)^4}
\frac{\mathcal{T}_a^{\mu\nu} \left(g_{ab}-\frac{q_a q_b}{q^2} \right)
	\gamma_{\mu} (\cancel{l}+m) \gamma_{\nu}}{l_1^2 l_2^2 (l^2-m^2)} \Big] u(k_1).
\end{align}

Any arbitrary expression can be decomposed onto a Dirac matrix basis,
\begin{align}
A = a_0 I + a_5 \gamma_5 + v_\mu \gamma^\mu + a_\mu \gamma_5 \gamma^\mu+ T_{\mu\nu}\sigma^{\mu\nu},
\end{align}
where coefficients determined by
\begin{equation}
\begin{split}
a_0 = \frac{1}{4} \mathrm{Tr}_D A, \quad a_5 = \frac{1}{4} \mathrm{Tr}_D (\gamma^5 A), \quad v^\mu = \frac{1}{4} \mathrm{Tr}_D (\gamma^\mu A), \\
a_\mu = - \frac{1}{4} \mathrm{Tr}_D (\gamma^5\gamma^\mu A), \quad T^{\mu\nu} = - \frac{1}{8} \mathrm{Tr}_D (\sigma^{\mu\nu}A).
\end{split}
\end{equation}
The only nonzero projection for axial-vector vertex is $a_\mu$.
As result one could write the meson-lepton effective interaction as $g_{eA}(Q^2) \bar{u}(k') \gamma^5 \gamma_{\mu} u(k)$, where
\begin{align}
g_{eA}(Q^2) &= i e^4 \int \frac{d^4 l}{(2\pi)^4} \frac{N}{l_1^2 l_2^2(l^2-m^2)}, \label{eq:g_eA_integral}
\\
N &= 2 \Big( \frac{(l\cdot q)}{q^2} \nu (l_1^2-l_2^2) F^{(0)}(Q^2,l_1^2,l_2^2)
+ (l\cdot l_1) \frac{X}{\nu} 
{F^{(1)}}(Q^2,l_2^2,l_1^2) + (l\cdot l_2)\frac{X}{\nu} F^{(1)}(Q^2,l_1^2,l_2^2) \Big). \label{eq:numerator_N}
\end{align}
%

To proceed one needs to know form factors $F^{(i)}$.
At present we have only few experimental data from L3 Collaboration on the  transition form factor \cite{L3_f1,L3_2} for $f_1(1285)$.
In this work the same parameterization as in \cite{Dorokhov:2017nzk} is used.
Namely, with the L3 data
one can parameterize $F^{(0)}$ as
\begin{align}
F^{(0)}(Q^2,l_1^2,l_2^2) =& F^{(0)}(Q^2,0,0) F(l_1^2) F(l_2^2), \\
F (l_i^2) =& \left(\frac{\Lambda^2}{\Lambda^2-l_i^2}\right)^2.
\end{align}
For the numerical estimate we fix the slope of form factors according to the L3 data to $\Lambda_{f_1(1285)}=1.04$~GeV.
Effect of off-shellness is considered by
\begin{equation}
F^{(0)}(Q^2,0,0)=F^{(0)}(M_A,0,0) e^{-(Q^2+M_A^2)/M_A^2}.
\end{equation}
The values of the form factors can be fixed from the L3 data using the relations given by the nonrelativistic quark model calculating the triangle anomaly diagram:
\begin{equation}\label{eq:A-gg_decay}
F^{(0)}(M_A^2,0,0)=-F^{(1)}(M_A^2,0,0), \qquad \tilde{\Gamma}_{\gamma^*\gamma^*} = \frac{\pi \alpha^2 M_A^5}{12} [ F^{(1)}(M_A^2,0,0)]^2,
\end{equation}
where $\tilde{\Gamma}_{\gamma^*\gamma^*}$ is a meson decay width.
Moreover, the nonrelativistic quark model requires the sign of $F^{(0)}$ is positive and
\begin{equation}\label{eq:F0_at_Mf1}
F^{(0)}_{f_1(1285)}(M_{f_1}^2,0,0) = (0.266 \pm 0.043)~\text{GeV}^{-2}
\end{equation}

If we use the result of nonrelativistic quark model that $F^{(0)}=-F^{(1)}$, then second and third parts of Eq.~(\ref{eq:numerator_N}) cancel out. We proceed assuming that contribution of $F^{(1)}$ part is small.

\begin{figure}
	\includegraphics[width=0.42\linewidth]{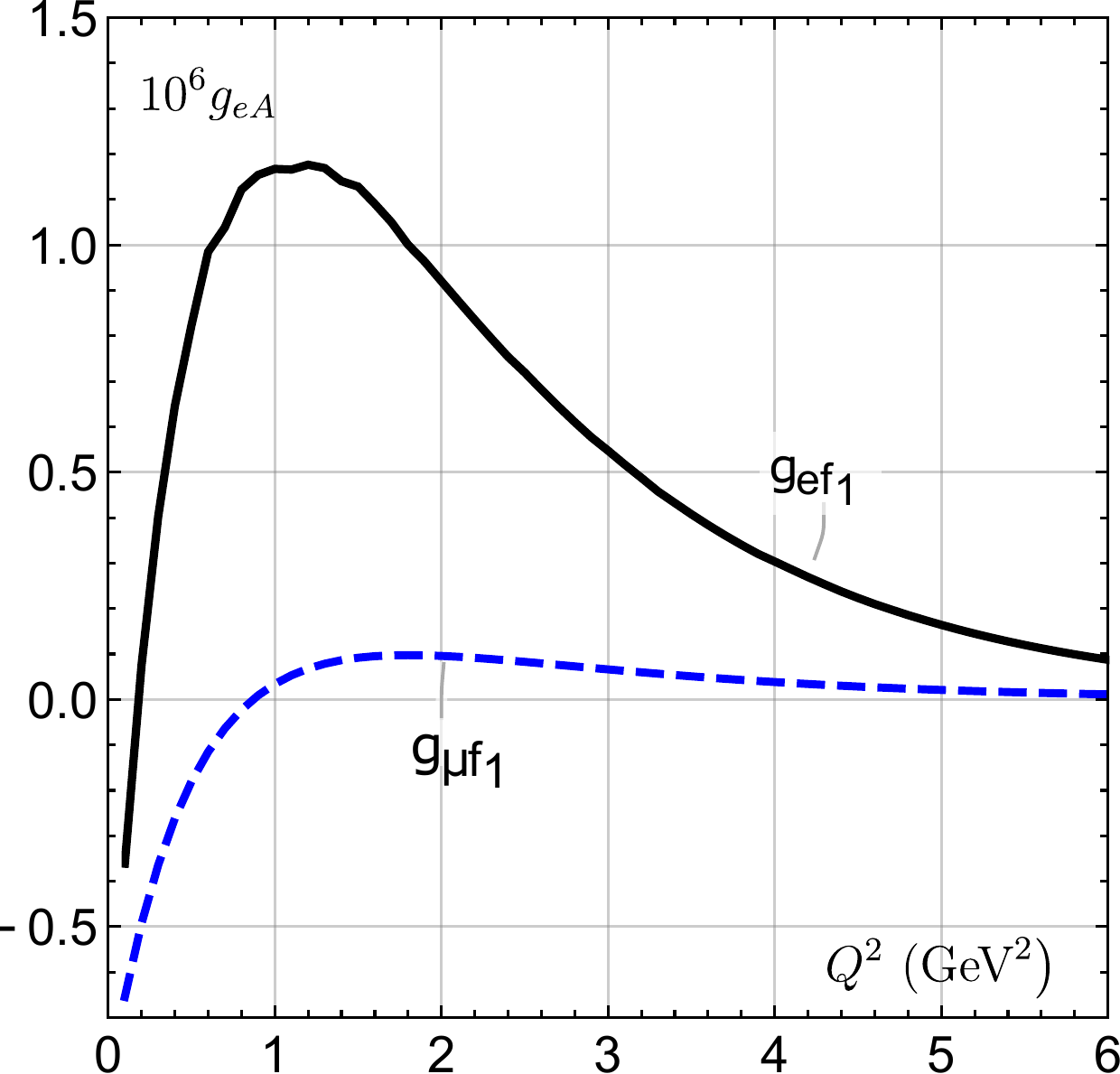} \quad \includegraphics[width=0.4\linewidth]{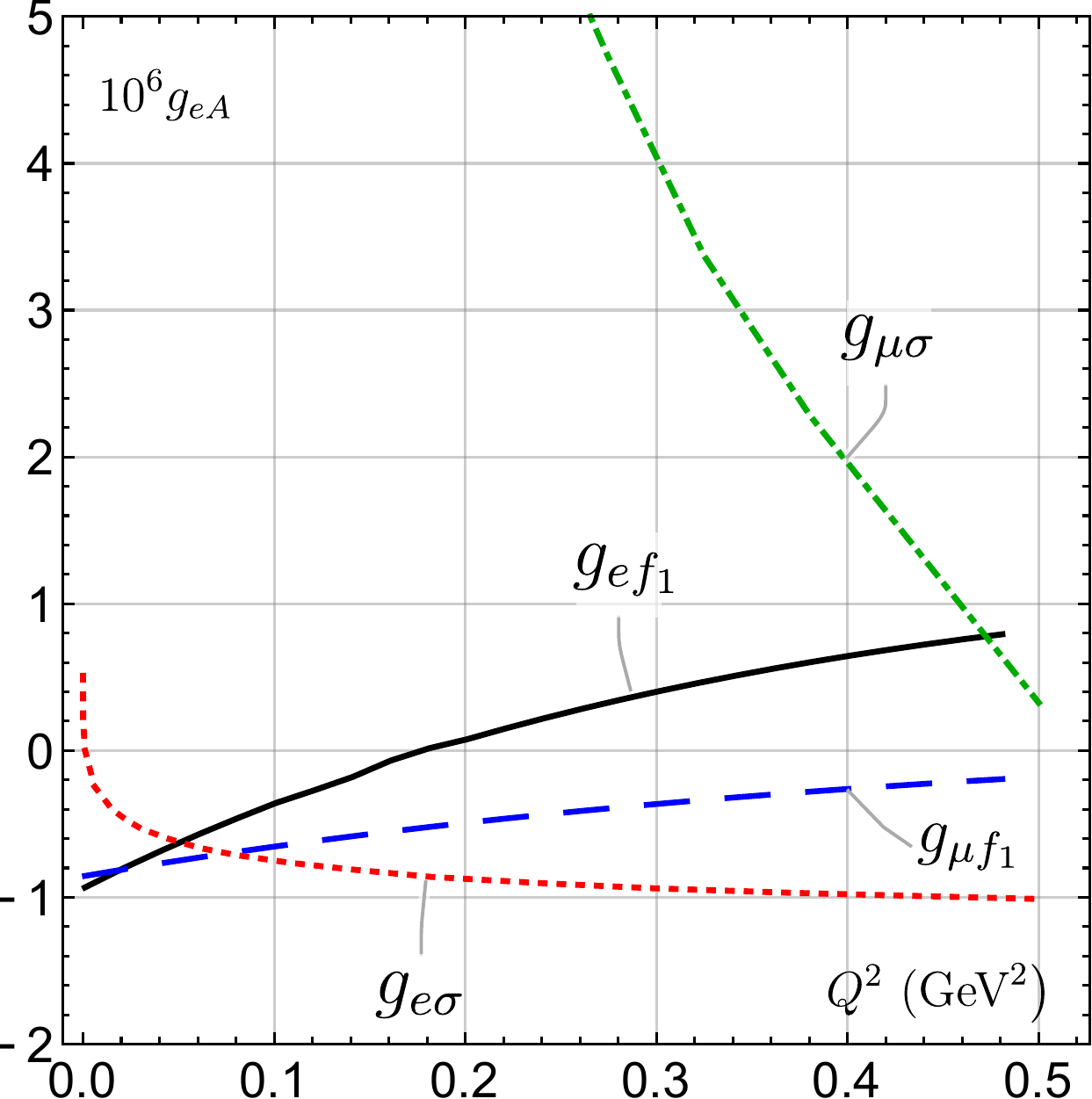}
	\caption{The effective coupling of axial $f_1(1285)$ meson to electron $g_{e f_1}$ and muon $g_{\mu f_1}$ as function of $Q^2$. Coupling of $\sigma$ meson from \cite{KoshchiiAfanasev:2016} is shown for comparison.}
	\label{fig:g_eA_plot}
\end{figure}

Now the integration of Eq.~(\ref{eq:g_eA_integral}) could be performed.
Details are given in Appendix \ref{Appemdix:geA}.
Final result for effective coupling of $f_1(1285)$ meson is shown in Fig.~\ref{fig:g_eA_plot} together with the result of $\sigma$ meson coupling from \cite{KoshchiiAfanasev:2016} for comparison.
We also calculated the coupling to muon, which turns out to be smaller.


The coupling to electron at very low $Q^2$ is same order of magnitude as $\sigma$ contribution in \cite{KoshchiiAfanasev:2016}.
Therefore it is important for precise low momenta scattering experiments for extraction of proton radius.
Contribution of others axial mesons as $f_1(1420)$ and $a_1(1260)$ summed up and total effect will be more significant.
Notice that in our calculation we neglect masses of external leptons, and our results for low $Q^2$ region should be treated as an estimation.

The last required unknown number is coupling to proton $g_{pA}$. 
It was estimated in \cite{Kirchbach:1995ep}  as $g_{f_1 N} = -(14.6 \div 17.2)$.
Substituting all numbers into Eq.~(\ref{eq:AV_correction_to_xsection}) and (\ref{eq:AV_polarization_correction_result}) one gets the final full result. In the next section it will be compared with OPE approximation.

\section{Results and discussion}\label{sec:results}

Rosenbluth method is based on the linearity of the reduced cross section in OPE with $\epsilon$. At $\epsilon=0$, $\sigma_R=\tau G_M^2$, and at  $\epsilon=1$, $\sigma_R=\tau G_M^2+G_E^2$. 
It is necessary to check how meson exchange contribution violates this linear dependency.
From Eq.~(\ref{eq:AV_correction_to_xsection}) one can see that at $\epsilon=0$ contribution is biggest and disappears at $\epsilon=1$.

TPE correction $\delta_{\gamma\gamma}$ to OPE cross section commonly presented as
\begin{equation}\label{eq:delta_TPE_deffinition}
\frac{d\sigma}{d\Omega} = \frac{d\sigma_{OPE}}{d\Omega} (1+\delta_{\gamma \gamma}).
\end{equation}
%
\begin{figure}[tb]
	\centering
	\includegraphics[width=0.49\linewidth]{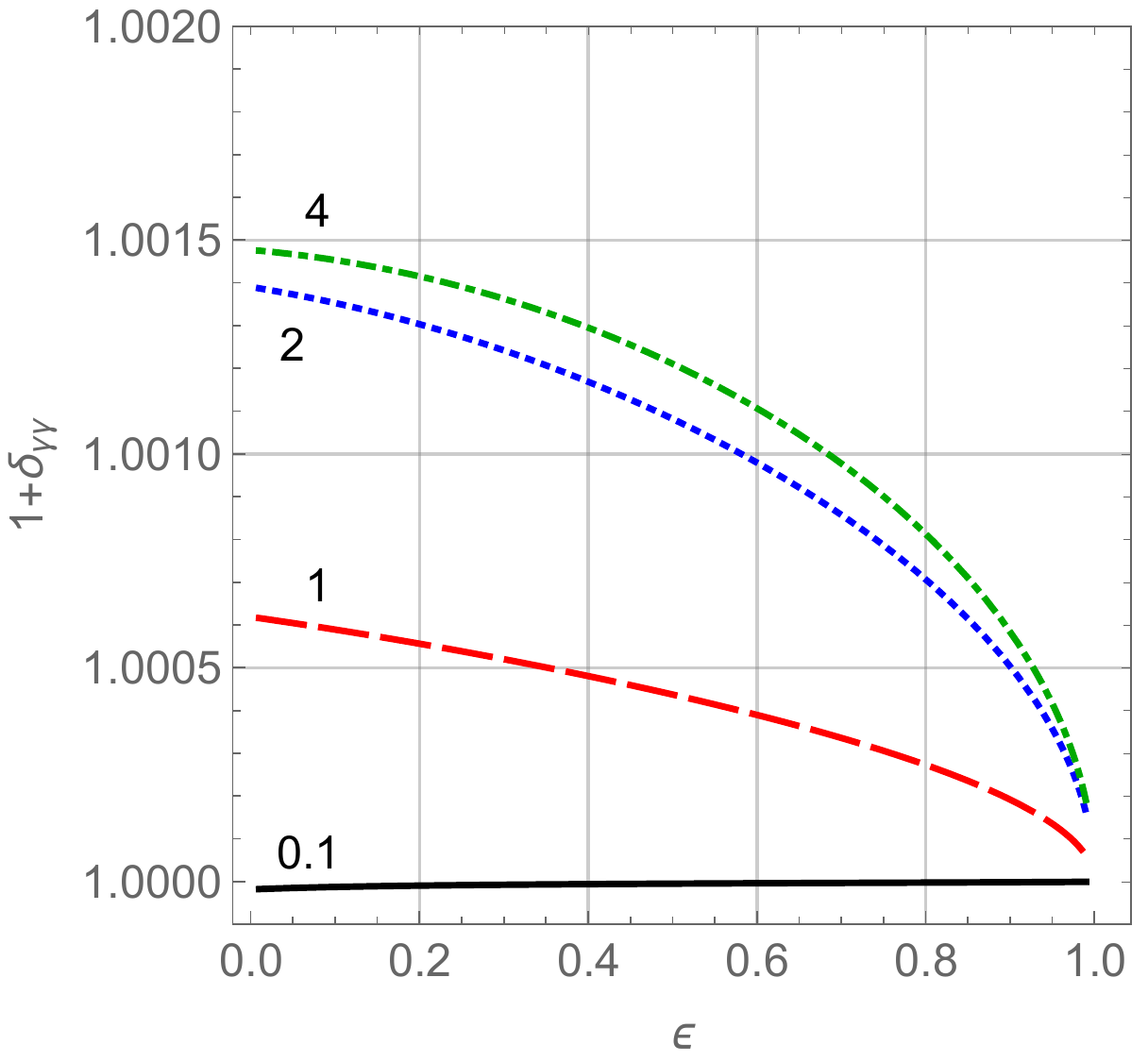}
	\includegraphics[width=0.49\linewidth]{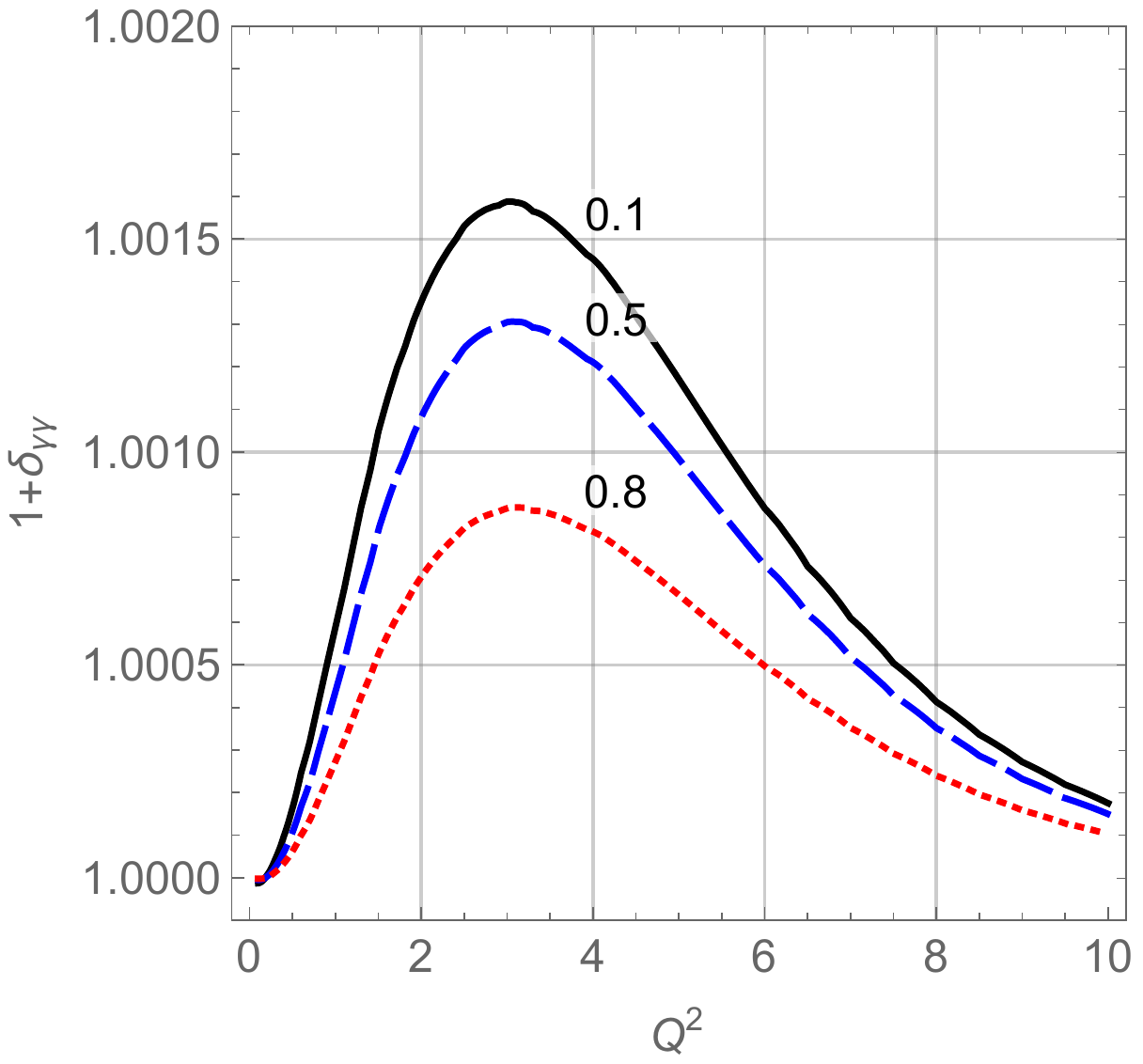}
	\caption{Left: TPE correction to OPE cross section  $(1+\delta_{\gamma\gamma})$ from Eq.~\eqref{eq:delta_TPE_deffinition} as function of $\varepsilon$ at different $Q^2$. Right: same as function of $Q^2$ at different $\varepsilon$.}
	\label{fig:delta_TPE}
\end{figure}
Fig.~\ref{fig:delta_TPE} demonstrates the correction produced by $f_1(1285)$ exchange.
One can see that it has maximum at $Q^2=3~\text{GeV}^2$.
Correction to OPE is of order of $0.1\%$
Box-like diagrams typically give correction on the level of few percents.
Our result is significantly smaller and can not affect extraction of form factors by Rosenbluth method.

In polarization transfer experiments ratio of form factors in OPE approximation is proportional to ratio of transverse to longitudinal polarization:
\begin{equation}\label{eq:R_ratio_PtPl}
R_{OPE} = \mu_p \frac{G_E}{G_M} = -\mu_p \sqrt{\frac{\tau (1+\varepsilon)}{2\varepsilon}} \frac{P_T}{P_L}.
\end{equation}
%
\begin{figure}[tb]
	\centering
		\includegraphics[scale=0.58]{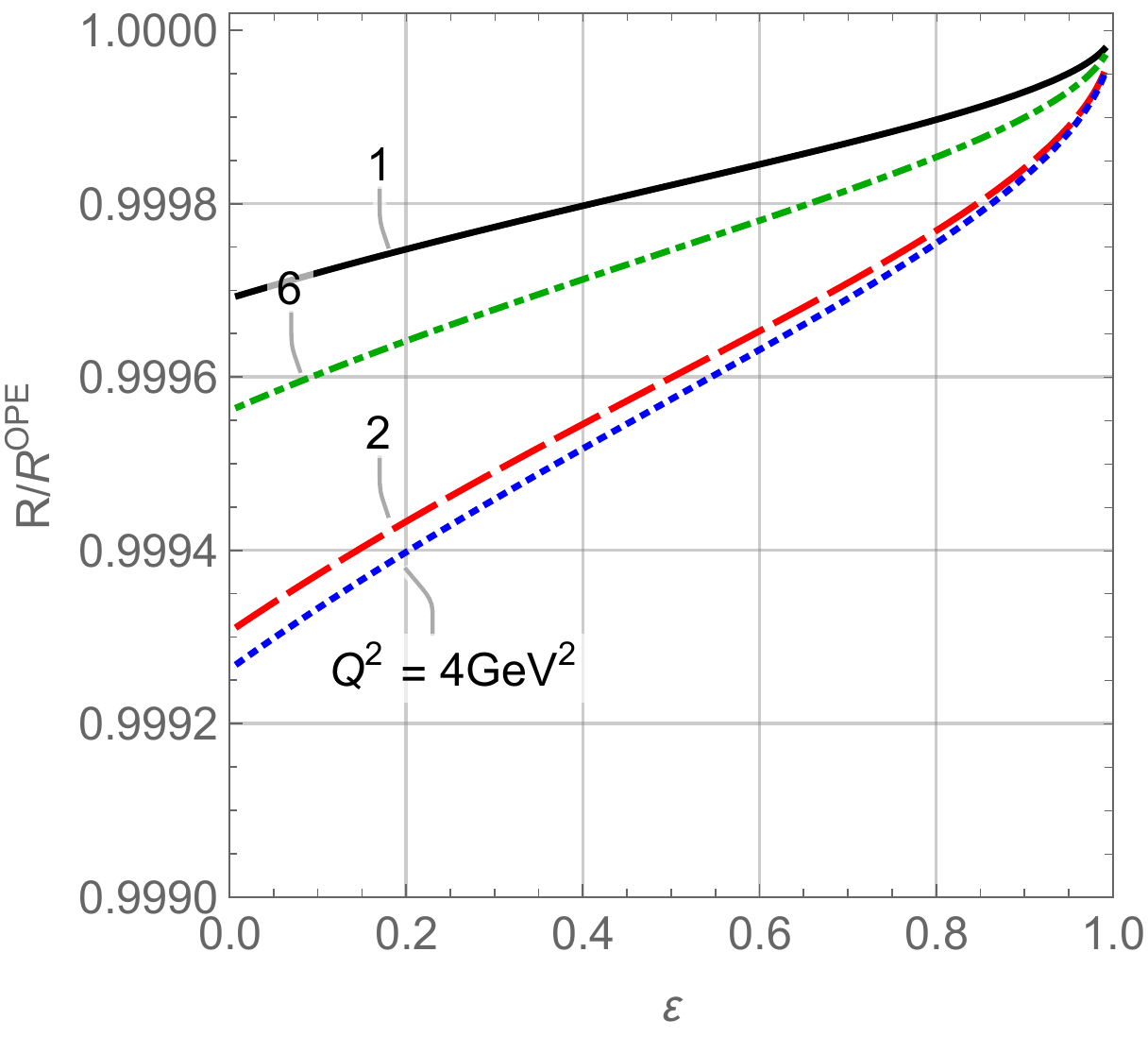} \quad
	\includegraphics[scale=0.58]{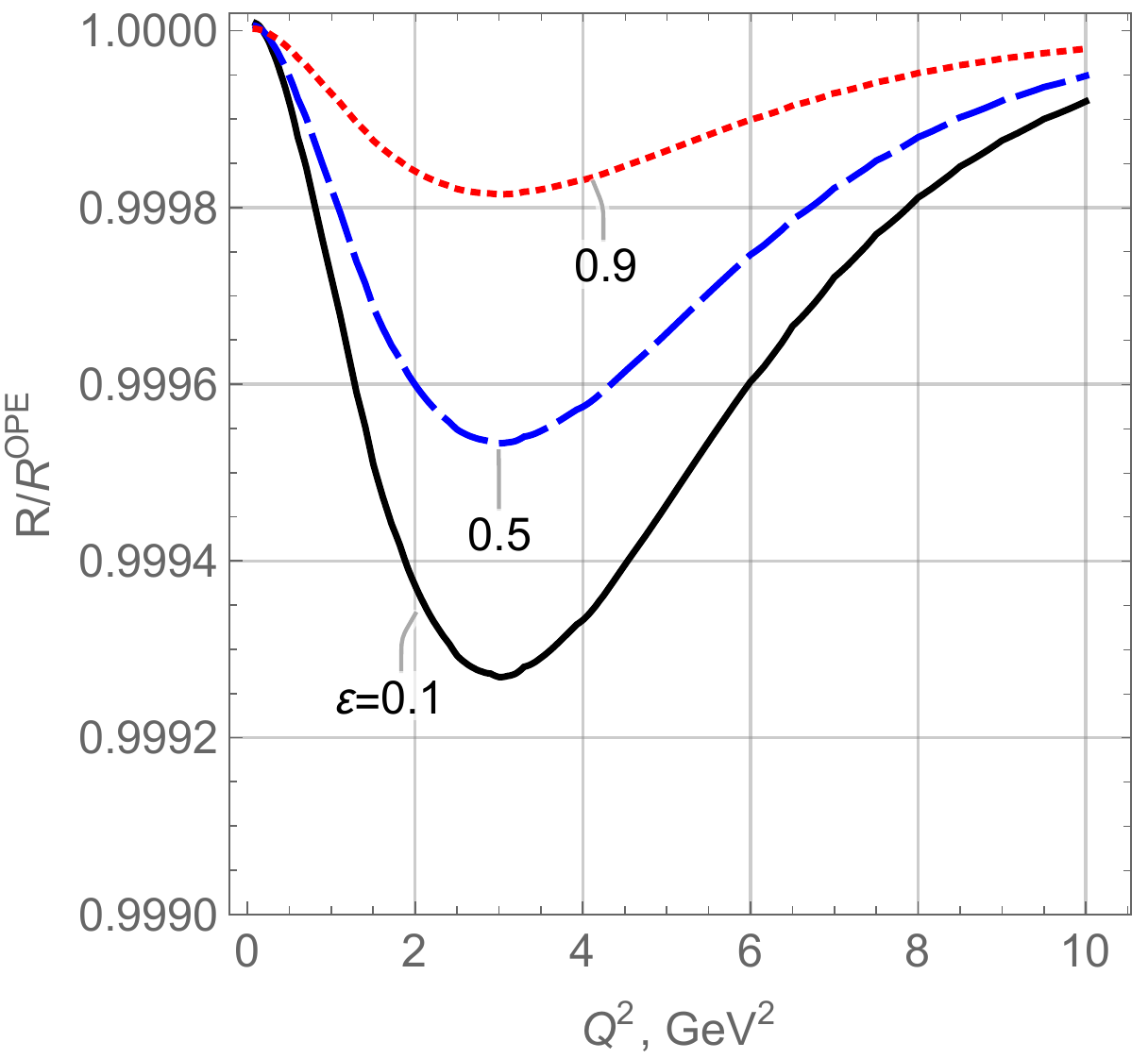}
	\caption{TPE correction to the ratio of transverse and longitudinal polarizations $R$. Right: as function of $Q^2$ at different $\varepsilon$, left: as function of $\varepsilon$ at different $Q^2$.}
	\label{fig:R_ratio_correction}
\end{figure}
Correction to this ratio can be represented as $R/R_{OPE}$, where $R$ is a ratio of polarizations calculated considering meson exchange.
This correction  is shown in Fig.~\ref{fig:R_ratio_correction}.
It is of order of $0.05\%$, which is not enough to influence form factors obtained from experimental data.
Finally, one can conclude that axial meson exchange could not affect form factors extracted by Rosenbluth and polarization transfer techniques.
\begin{figure}[t]
	\centering
	\includegraphics[scale=0.58]{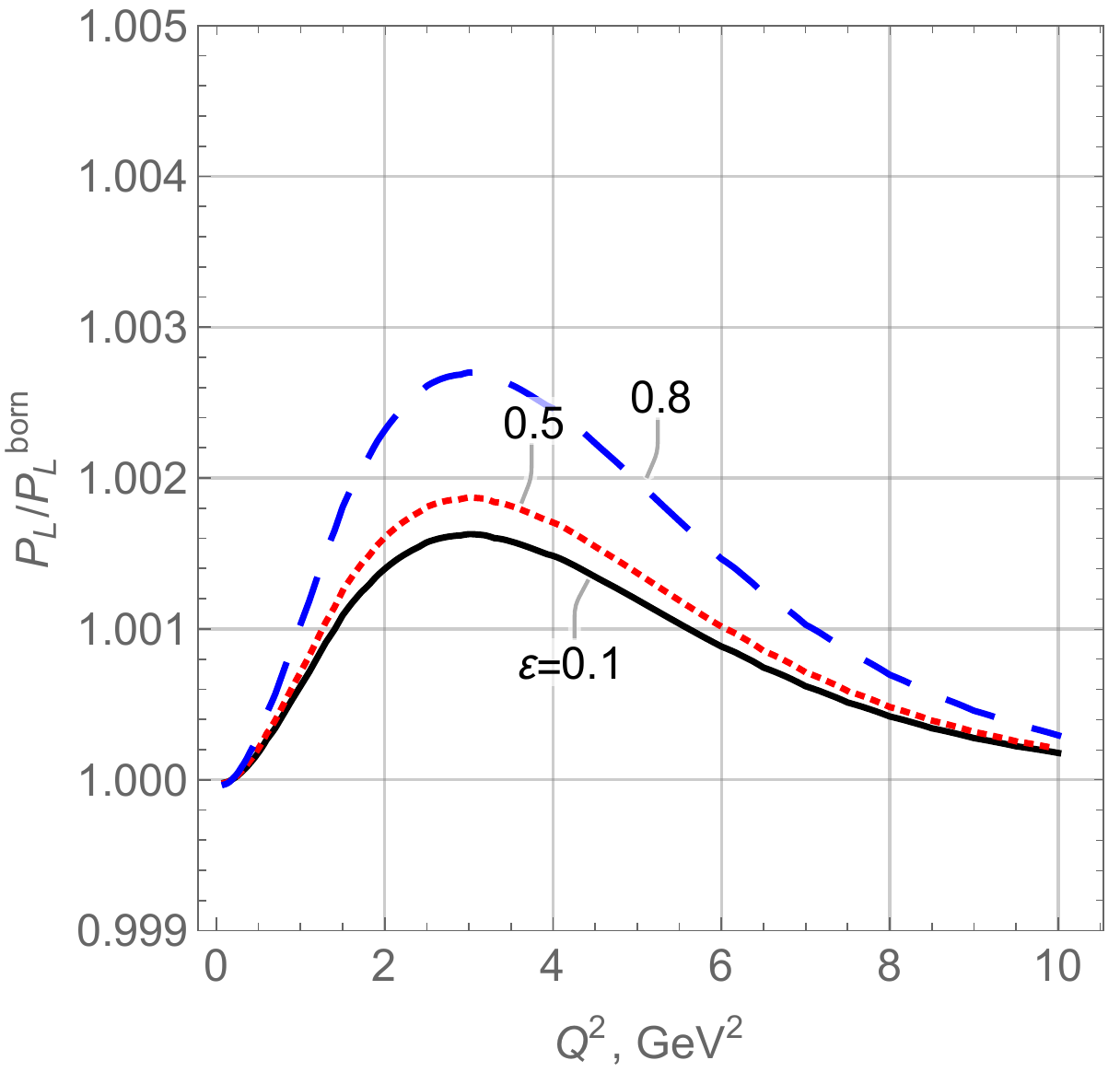} \quad
	\includegraphics[scale=0.58]{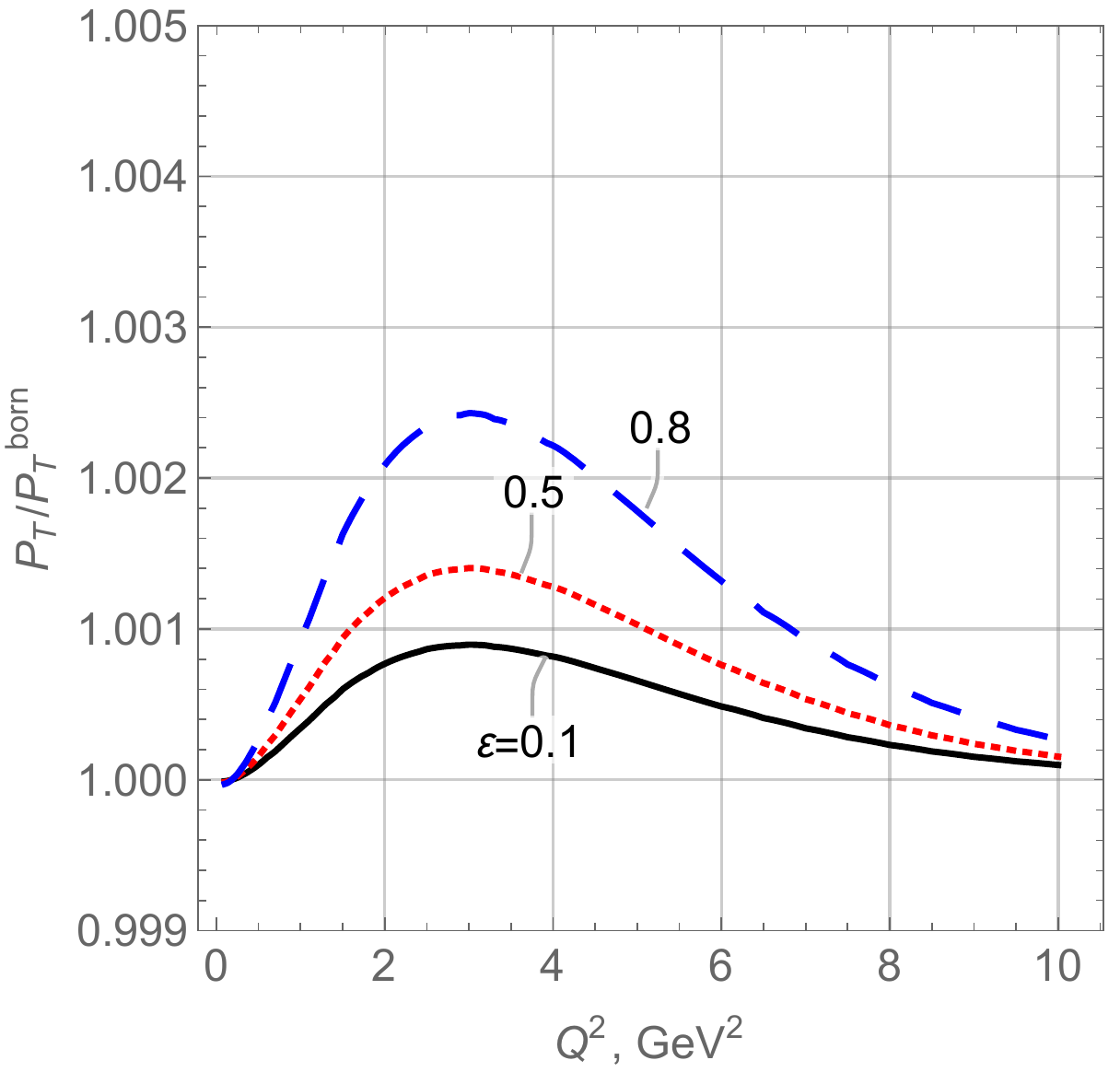} \\
	\includegraphics[scale=0.58]{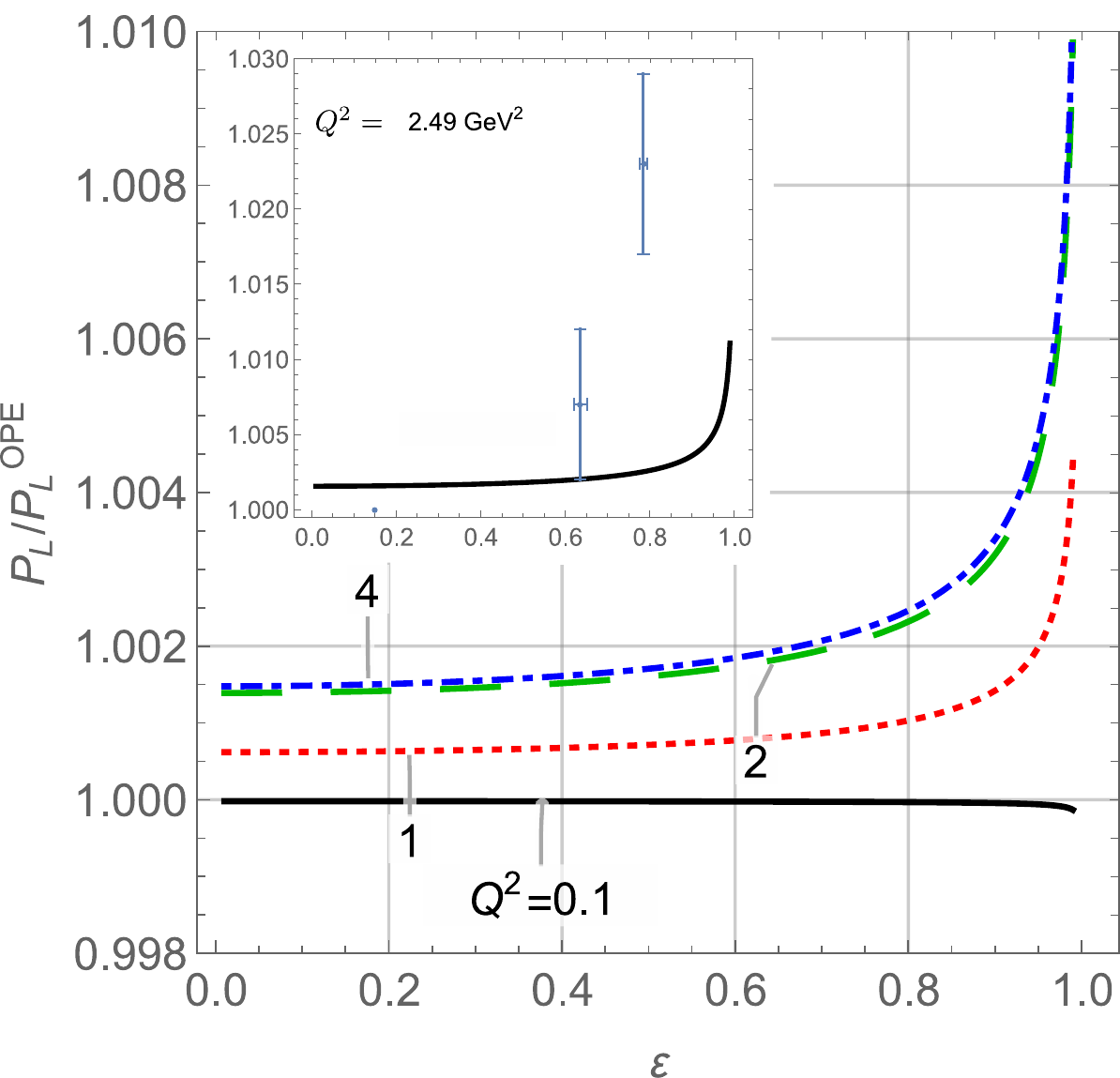} \quad
	\includegraphics[scale=0.58]{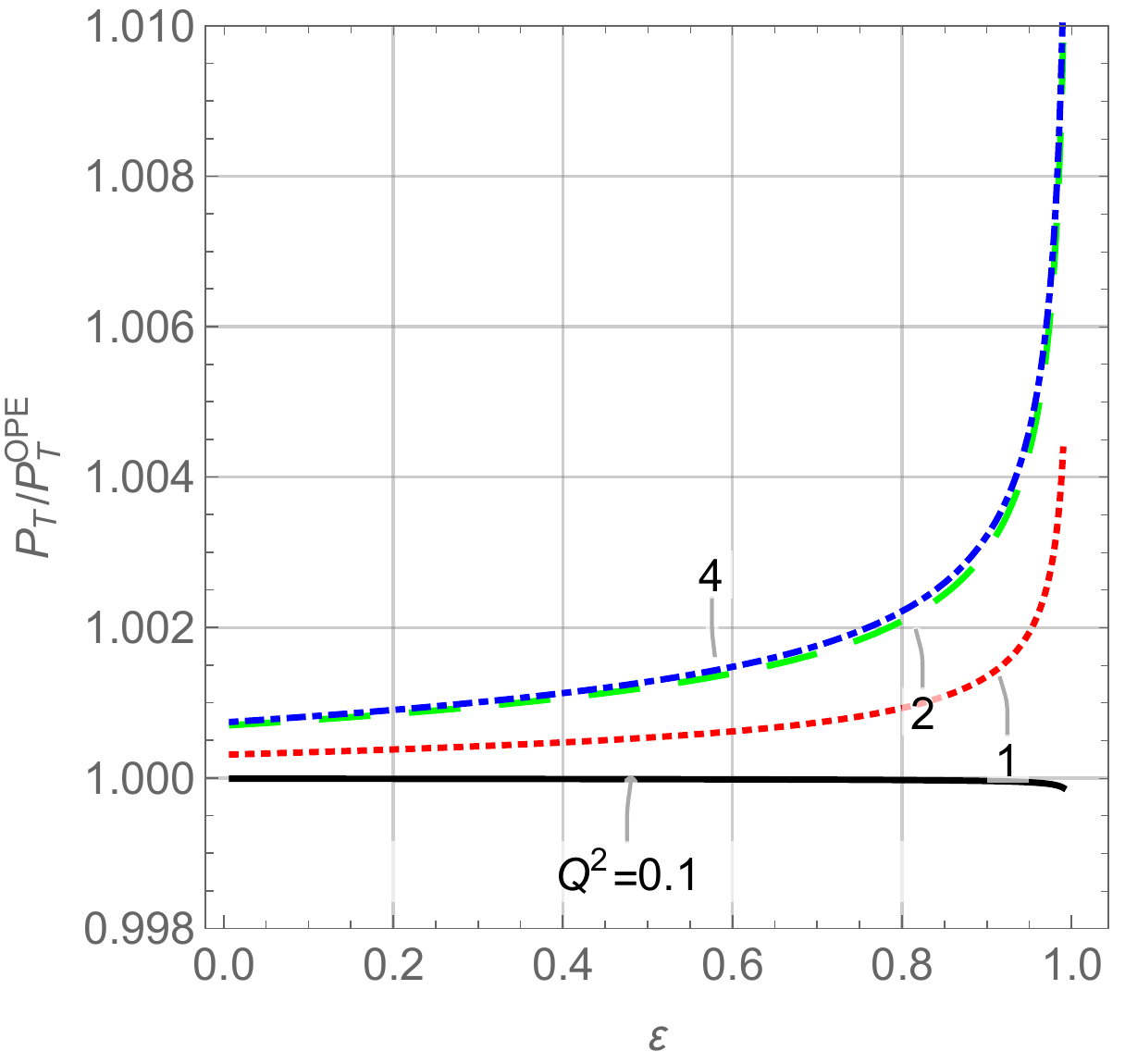}

	\caption{TPE correction to OPE polarization transfer. Left: longitudinal, right: transverse.  
	Data for longitudinal polarization are from  
	\cite{GEp2gammaCollMeziane:2010},\cite{Puckett:2017flj}(only statistical error).}
\label{fig:correction_to_PL_PT}
\end{figure}

Deviation of individual polarizations is also subject of interest since longitudinal polarization was measured experimentally in \cite{GEp2gammaCollMeziane:2010}.
Corrections are depicted in Fig.~\ref{fig:correction_to_PL_PT}.
It also shows comparison with experimental points for longitudinal polarization at $Q^2=2.49~\GeV^2$.
One can see that result is close to data points.
Contributions of other mesons will add up rising the theoretical line and could explain data. However it requires more detailed analysis.


One can notice from Eq.~\eqref{eq:AV_correction_to_xsection} and Eq.~\eqref{eq:AV_polarization_correction_result} that corrections to cross section depend on form factor itself. 
The sensitivity of results is checked by using different form factors. 
The comparison with dipole form factor and Kelly fit\cite{Kelly:2004hm} is performed and difference is negligible.
\begin{figure}[tbh]
	\centering
	\includegraphics[scale=0.58]{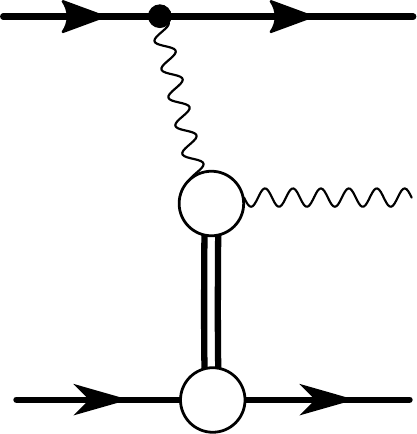}
	\caption{Possible mechanism for real photon production from meson exchange which can affect bremsstrahlung.}
	\label{fig:real_gamma_production}
\end{figure}

The axial meson exchange could not be responsible for  difference of form factors extracted by Rosenbluth and polarization transfer techniques. However it also could affect extraction indirectly by contribution to final state radiation.
It is interesting to consider a diagram shown in Fig.~\ref{fig:real_gamma_production}. 
It is not suppressed by additional QED constant in comparison with diagrams considered in this work. However it supressed by aditional $\alpha_{QED}$ in respect to usual radiative corrections.

\section{Conclusion}


Contribution of axial-vector meson exchange to unpolarized $ep$ cross section and polarization transfer is calculated.
For this purpose, effective lepton-meson coupling was estimated.

The coupling of electron to $f_1(1285)$ is same order of magnitude as to scalar  $\sigma$ meson: $g_{ef_1} \approx 10^{-6}$.
It changes sign at very low $Q^2\approx 0.2~\GeV^2$ and has maximum at $1~\GeV^2$. 
Meanwhile, coupling $f_1(1285)$-muon is much smaller.


Correction to OPE for unpolarized cross section is of order $0.1\%$, which is significantly smaller than the current experimental precision. 
Correction has the maximum at $\varepsilon=0$ and $Q^2\approx 3~\GeV$.
For polarization ratio correction has the similar scale: $0.05\%$. 
More interesting result is for individual polarizations.
It has correct qualitative behavior and decrease difference  between theory and experiment for longitudinal polarization. Consideration of other mesons could explain data.


Finally, one can conclude that contributions of axial mesons exchanges could not be responsible for difference of form factors extracted by Rosenbluth and polarization transfer methods. However results support the idea that meson exchange could be behind deviation from OPE of individual polarizations.

\begin{acknowledgments}
	Authors thank [N.I. Kochelev], a great mentor and scientist, who proposed the idea of this work.
	Authors also thank A.A. Osipov for useful discussions.
	The study was supported by the National Natural Science Foundation of China, No.11875296 and the Chinese Academy of Sciences President’s International Fellowship Initiative via Grants No. 2020PM0073.
\end{acknowledgments}

\appendix

\section{Calculation of $g_{eA}$}\label{Appemdix:geA}

This appendix is dedicated to details of lepton-meson coupling calculation. $g_{eA}$ is the product of the following integral
\begin{align}
g_{eA} =& i e^4 \int \frac{d^4 l}{(2\pi)^4} \frac{N}{l_1^2 l_2^2(l^2-m^2)}, \label{eq:g_eA_integral_appendix}
\\
N =& 2 \Big( \frac{((k+l_1)\cdot q)}{q^2} \nu (l_1^2-l_2^2) F^{(0)}(l_1^2,l_2^2) \nonumber
\\
&+ ((k+l_1)\cdot l_1) \frac{X}{\nu} {F^{(1)}}(l_2^2,l_1^2) + ((k+l_1)\cdot l_2)\frac{X}{\nu} F^{(1)}(l_1^2,l_2^2) \Big). \label{eq:numerator_N_appendix}
\end{align}
%


In the NRQM $F^{(0,1)}$ form factors are related with Adler's form factors as
\begin{align}
F^{(0)}(l_1^2,l_2^2)= - A_4, \quad F^{(1)}(l_1^2,l_2^2)=\frac{\nu}{X} (l_2^2-\nu) A_4, \quad {F^{(1)}}(l_2^2,l_1^2) = \frac{\nu}{X} (l_1^2-\nu) A_4.
\end{align}
In such case terms with $F^{(1)}$ cancels out.  
Therefore one can assume that the main contribution comes from $F^{(0)}$ term.
Thus considering it and rewriting  Eq.~\eqref{eq:numerator_N_appendix} in terms of momenta shown in Fig.\ref{fig:loop}, the numerator $N$ becomes
\begin{align}
N =& \frac{2}{q^2} (l\cdot p - l^2 -k_1 \cdot k_2) (l \cdot q)^2 F^{(0)},
\end{align}
where $p=k_1 + k_2$ and $p\cdot q=0$.
Form factor is parameterized as
\begin{align}
F^{(0)}(l_1^2,l_2^2) =
\left(\frac{\Lambda^2}{\Lambda^2-l_1^2}\right)^2
\left(\frac{\Lambda^2}{\Lambda^2-l_2^2}\right)^2.
\end{align}
The full expression for lepton-meson coupling is
\begin{align}\label{eq:g_eA_integral}
	g_{eA}(Q^2) =& i e^4
	F^{(0)}(M_A^2,0,0)  e^{-(Q^2+M_A^2)/M_A^2} \times 
	\nonumber \\
	&\int \frac{d^4 l}{(2\pi)^4} \frac{2 (l\cdot p - l^2 -k_1 \cdot k_2) (l \cdot q)^2}{q^2 (l^2-m^2)(l-k_1)^2 (l-k_2)^2}
\left(\frac{\Lambda^2}{\Lambda^2-l_1^2}\right)^2
\left(\frac{\Lambda^2}{\Lambda^2-l_2^2}\right)^2.
\end{align}
Note that Eq.~(\ref{eq:numerator_N_appendix}) is for a meson on the mass-shell. The effect of off-shellness is taken into account by the factor $\exp(-(Q^2+M_A^2)/M_A^2)$. $F^{(0)}(M_A^2,0,0)$ is form factor normalization.

The integral could be calculated using following general scheme: use Laplace transformation, do shift in integration momenta $l$ to make integral Gaussian and take it.
Then, integral over $\alpha$-s is transformed to relative coordinates $L$ and $x_i$.
$L$ integration is also Gaussian. Multidimensional integral over $x_i$ is done numerically.

First step is to apply following integral representation to propagators
\begin{equation}
	\frac{1}{(p_i^2-m^2 + i\varepsilon)^k} = \frac{(-i)^k}{\Gamma(k)} \int_0^{\infty} d\alpha_i e^{i \alpha_i (p_i^2-m^2 +i \varepsilon)} \alpha_i^{k-1}.
\end{equation}
As result one gets the exponent in power of
\begin{align}
\alpha_1 l_1^2 + \alpha_2 l_2^2 + \alpha_3 l^2 + \alpha_4 l_1^2 + \alpha_5 l_2^2 + B
&= \bar{a}_1 l_1^2 + \bar{a}_2 l_2^2  + \alpha_3 l^2 + B  \\
&= \Delta l^2 + l \cdot (\bar{a}_1 k_1 + \bar{a}_2 k_2) + B,
\end{align}
\begin{align}
\bar{a}_1 = \alpha_1 + \alpha_4; \qquad \bar{a}_2 = \alpha_2 + \alpha_5, \qquad
B = -\alpha_3 m^2 - (\alpha_4 + \alpha_5) \Lambda^2 , \qquad
\Delta = \sum_i \alpha_i,&
\end{align}
where $B$ denotes the part independent of $l$. 
Transition to integral representation adds $(-i)^7 = i$ as a common factor.

Further, one needs to do momentum shift and take Gaussian integral over $l$. Shifting $l \to l - V$ by vector $V = \frac{\bar{a}_1 k_1 + \bar{a}_2 k_2}{\Delta}$, keeping even power of $l$ and applying $l^\mu l^\nu = g^{\mu\nu} l^2/4$, numerator becomes
\begin{align}
\left(
\frac{l^2 q^2}{4} + (V \cdot q)^2
\right)
\left(
V \cdot p - l^2 - V^2 + \frac{q^2}{2}
\right)
- l^2 (V \cdot q)^2.
\end{align}
After shifting the expression in exponent becomes
\begin{align}
\Delta l^2 + \frac{1}{\Delta}(\bar{a}_1 \bar{a}_2 q^2 + \bar{a}_1 \alpha_3 k_1^2 + \bar{a}_2 \alpha_3 k_2^2) + B.
\end{align}
In massless limit of initial and final leptons it simplified to
\begin{align}
\Delta l^2 + \frac{1}{\Delta}(\bar{a}_1 \bar{a}_2 q^2) + B.
\end{align}
Full result is proportional to
\begin{align}
\propto \int \!\!\! d^4l \, \frac{2}{q^2} \Bigg[
\left(\frac{l^2 q^2}{4} + (V \cdot q)^2\right)
\left(V \cdot p - l^2 - V^2 + \frac{q^2}{2} \right)
- l^2 (V \cdot q)^2
\Bigg]
e^{i \Delta l^2}
\exp \Big[\frac{i}{\Delta}(\bar{a}_1 \bar{a}_2 q^2 + B) \Big].
\end{align}
After factoring out same powers of $l$ one gets three terms:
\begin{align}
\propto l^0: & \quad I_1 = \frac{2}{q^2} (V \cdot q)^2  (  V \cdot p - V^2 + \frac{q^2}{2}), \\
\propto l^2: & \quad I_2 = \frac{1}{2} ( V \cdot p  - V^2 + \frac{q^2}{2}) - \frac{4}{q^2} (V \cdot q)^2, \\
\propto l^4 : & \quad I_3 =  -\frac{1}{2}.
\end{align}
Integration over $l$ can be performed after Wick rotation $l^0=i l^0_E$ and for every term gives
\begin{align}
I_1 \int d^4 l e^{i \Delta l^2} = I_1 \frac{- i \pi^2}{\Delta^2}, \\
I_2 \int d^4 l \, l^2 e^{i \Delta l^2} =  I_2 \frac{2 \pi^2}{\Delta^3},\\
I_3 \int d^4 l \, l^4 e^{i \Delta l^2} = I_3 \frac{ i 6 \pi^2}{\Delta^4},
\end{align}

Next step is to transform integrals over $\alpha$ using standard technique
\begin{align}
\int_0^\infty \prod_i d \alpha_i &= \int dL L^4 \prod_i d x_i \delta(1-\sum_i x_i)
\\
&= \int\displaylimits_0^\infty L^4 d L \int\displaylimits_0^1 \! d x_1 \int\displaylimits_0^{1-x_1} \! dx_2 \hspace{-0.6em} \int\displaylimits_0^{1-x_1-x_2} \hspace{-1em} dx_3 \hspace{-0.3em} \int\displaylimits_0^{1-x_1-x_2-x_3} \hspace{-1em} dx_4,
\end{align}
\begin{align}
\Delta = L, \qquad \alpha_i = x_i L.
\end{align}
In such case expressions modify as following
\begin{align}
&V = \bar{x}_1 k_1 + \bar{x}_2 k_2, &
&\bar{x}_1 = x_1 + x_4,  &
&\bar{x}_2 = x_2 + x_5, \qquad \\
&V^2 = - q^2 \bar{x}_1 \bar{x}_2,&
&(V \cdot q)^2 = \frac{q^4}{4} (\bar{x}_2-\bar{x}_1)^2, &
&(V \cdot p) = -\frac{q^2}{2} (\bar{x}_1 + \bar{x}_2),
\end{align}
\begin{align}
[ V \cdot p - V^2 + \frac{q^2}{2}] &= \frac{q^2}{2}[1 - \bar{x}_1 - \bar{x}_2 + 2 \bar{x}_1 \bar{x}_2], \\
I_1 &= \frac{q^4}{4} (\bar{x}_2-\bar{x}_1)^2 [1 - \bar{x}_1 - \bar{x}_2 + 2 \bar{x}_1 \bar{x}_2], \\
I_2 &= \frac{q^2}{4}[1 - \bar{x}_1 - \bar{x}_2 + 2 \bar{x}_1 \bar{x}_2] - q^2 (\bar{x}_2-\bar{x}_1)^2.
\end{align}
Now one has to calculate the sum of three convergent integrals over $L$:
\begin{align}
-\frac{e^4  \Lambda^8}{(2\pi)^4} x_4 x_5\left(I_1 (-i) \pi^2 \int dL e^{i L A} L^4 + I_2 2 \pi^2 \int dL e^{i L A} L^3 
+ I_3 i 6 \pi^2 \int dL e^{i L A} L^2 \right),\\
A = \bar{x}_1 \bar{x}_2 q^2 - x_3 m^2 - (x_4+x_5)\Lambda^2.
\end{align}
This integrals is easy to calculate using
\begin{align}
\int_0^\infty \!\! dL \, e^{-i A L} L^n = (-i)^{n+1} \frac{n!}{A^{n+1}}.
\end{align}
The final result is three integrals: the first one is
\begin{align}
\frac{e^4 \Lambda^8 \pi^2}{(2\pi)^4} \int \{dx_i\} \frac{4! I_1 x_4 x_5 }{A^5}
=
6 \alpha_{\mathrm{QED}}^2 \Lambda^8 Q^4 \int \{dx_i\} \frac{x_4 x_5(\bar{x}_2-\bar{x}_1)^2 [1 - \bar{x}_1 - \bar{x}_2 + 2 \bar{x}_1 \bar{x}_2] }{(x_3 m^2 + (x_4+x_5)\Lambda^2 + Q^2 \bar{x}_1 \bar{x}_2)^5},
\end{align}
the second integral
\begin{align}
\frac{e^4 \Lambda^8 \pi^2}{(2\pi)^4} \int \{dx_i\} \frac{-2 \cdot 3! I_2 x_4 x_5}{A^4}
=
12 \alpha_{\mathrm{QED}}^2 \Lambda^8 Q^2 \int \{dx_i\} x_4 x_5\frac{1/4[1 - \bar{x}_1 - \bar{x}_2 + 2 \bar{x}_1 \bar{x}_2] - (\bar{x}_2-\bar{x}_1)^2 }{(x_3 m^2 + (x_4+x_5)\Lambda^2 + Q^2 \bar{x}_1 \bar{x}_2)^4},
\end{align}
and the third integral
\begin{align}
\frac{e^4 \Lambda^8 \pi^2}{(2\pi)^4} \int \{dx_i\} \frac{12 I_3 x_4 x_5}{A^3}
=
\alpha_{\mathrm{QED}}^2 \Lambda^8  \int \{dx_i\} \frac{-6 x_4 x_5}{(x_3 m^2 + (x_4+x_5)\Lambda^2 + Q^2 \bar{x}_1 \bar{x}_2)^3},
\end{align}
where $q^2=-Q^2$ was used. Coupling $g_{eA}$ proportionals to the sum of that three integrals.
Integration over $x$ was done numerically. 
Result is shown in Fig.\ref{fig:g_eA_plot}.


\end{document}